\documentclass[sigconf]{acmart}
\AtBeginDocument{%
  }

\copyrightyear{2026}
\acmYear{2026}
\setcopyright{cc}
\setcctype{by}
\acmConference[KDD '26]{Proceedings of the 32nd ACM SIGKDD Conference on Knowledge Discovery and Data Mining V.2}{August 09--13, 2026}{Jeju Island, Republic of Korea}
\acmBooktitle{Proceedings of the 32nd ACM SIGKDD Conference on Knowledge Discovery and Data Mining V.2 (KDD '26), August 09--13, 2026, Jeju Island, Republic of Korea}
\acmDOI{10.1145/3770855.3818461}
\acmISBN{979-8-4007-2259-2/2026/08}

\newcommand{\sectitle}[1]{\par\addvspace{\smallskipamount}\noindent\textbf{#1}}

\begin{document}

\title{PinEqualizer: Full Funnel Content Exploration and Debiasing System at Pinterest}

\settopmatter{authorsperrow=4}

\newcommand{\pinterestauthor}[2]{%
  \author{#1}%
  \email{#2@pinterest.com}%
  \affiliation{%
    \institution{Pinterest, Inc.}%
    \city{San Francisco}%
    \country{USA}%
  }%
}

\pinterestauthor{Olafur Gudmundsson}{ogudmundsson}
\pinterestauthor{Bo Zhao}{bozhao}
\pinterestauthor{Huayi Liao}{hliao}
\pinterestauthor{Anna Kiyantseva}{akiyantseva}
\pinterestauthor{Sai Xiao}{sxiao}
\pinterestauthor{Heath Vinicombe}{heathvinicombe}
\pinterestauthor{Mostafa Keikha}{mkeikha}
\pinterestauthor{Luke DeLuccia}{ldeluccia}
\pinterestauthor{Zihao Chen}{zichen}
\pinterestauthor{Junpeng Hou}{jhou}
\pinterestauthor{Weijie Jiang}{weijiejiang}
\pinterestauthor{Bhawna Juneja}{bjuneja}
\pinterestauthor{Andreanne Lemay}{alemay}
\pinterestauthor{Wei-Ting Lin}{wlin}

\author{Keyvan Moghadam}
\authornote{Work done at Pinterest.}
\email{keyvanrmo@gmail.com}
\affiliation{
  \institution{Pinterest Inc.}
  \city{San Francisco}
  \country{USA}
}

\pinterestauthor{Jiaxing Qu}{jqu}
\pinterestauthor{Zhiqing Rao}{zrao}
\pinterestauthor{Zhihua Zhang}{zzhang}

\renewcommand{\shortauthors}{Olafur Gudmundsson et al.}

\begin{abstract}
In this paper, we propose a new solution for addressing the content cold-start problem in industry-scale search and recommender systems. Compared to prior approaches, we have made the following new contributions: 1) our solution spans the entire multi-stage funnel and generalizes well for both search and recommendation surfaces, 2) our solution reduces bias favoring existing content, allowing more accurate model prediction across content types and reducing short-term tradeoffs associated with high volumes of explicit content exploration, 3) our solution is evaluated with a scalable measurement framework that enables fast short-term experimentation while validating long-term impact. We have iteratively built and successfully deployed this new system at Pinterest in the past two years and observed significant improvements in fresh content exploration, overall user engagement, and content ecosystem health.
\end{abstract}



\maketitle

\section{Introduction}
\label{sec:intro}
Maintaining an ecosystem of high-quality, engaging content is the lifeblood of all successful search and recommender systems (RecSys). Pinterest, a visual discovery engine of over half a billion users and hundreds of billions of unique content items ("Pins"), is no exception. Matching these users with content sourced from a constantly evolving corpus requires solving the "cold start" problem \cite{sigir02-coldstart}, in which content with low prior exposure struggles to gain traction.

The cold-start problem has been well studied, with one of the most effective solutions being leveraging content-based matching between users and items. In line with other industry scale platforms, Pinterest systems already heavily rely on both content-based signals and its unique pin-board graph structure. For example, graph random walk \cite{eksombatchai2018pixie} is used in candidate generation to fetch other related Pins saved on the same board, and graph-optimized content embeddings \cite{PinSage} and the corresponding user embeddings \cite{PinSage, PinnerFormer} and user activity sequence \cite{TransAct} are among the most effective signals in our ranking models. However, as content acquisition evolves over time, a significant amount of new content is no longer well-connected to the user-generated pin-board graph. Therefore, the cold-start problem becomes more serious over time and begins deteriorating the health of the content ecosystem: a rich-get-richer effect develops which advantages established content providers over new entrants.

In order to solve these challenges, we have built and deployed to production a new, end-to-end solution across the three main product surfaces at Pinterest: Homefeed, Related Pins, and Search, responsible for 92\% of total content impressions on the platform. Compared with previously published work, we have made contributions and innovations in the following areas:

\sectitle{Full-funnel approach:} Industry-scale RecSys are typically multi-stage systems. Our system innovates across the entire funnel from corpus selection, retrieval, to late stage ranking and utility. Our proposed solutions can generalize well across both search and recommendation surfaces, and we have also built search-specific optimizations that guarantee strong semantic relevance.
    
\sectitle{Debiasing existing content:} We found there is often bias in the entire funnel favoring existing content, and therefore it is important to reduce such bias instead of only focusing on exploration mechanisms. Debiasing allows models to predict more accurately across all content and as a result reduce the need for explicit exploration (via UCB or Thompson Sampling) and its corresponding short-term engagement tradeoff.

\sectitle{Scalable measurement framework:} We propose a comprehensive measurement framework. A content holdout experiment is deployed to continuously measure the long-term value of content exploration and debiasing improvements in production. In addition, we identified short-term experiment metrics that correlate with the long-term outcomes, but are measurable via user-segmented A/B experiments, significantly accelerating the speed of building and testing new improvements by engineering teams. Our setup is less complex and can successfully address several limitations in industry-scale production systems compared with previous work \cite{wsdm24-explore}.

We have iteratively built and successfully deployed various parts of the overall system since 2024 and collectively contributed to a 350\% increase in fresh content impressions with significant long term improvements in user engagement as well as content ecosystem health (e.g., content provider diversity) at Pinterest. We hope that the solutions and insights presented in this paper can provide value for RecSys practitioners interested in solving similar challenges.

\section{Measurement \& Metrics}
\label{sec:measurement}
In this work, we establish a holistic measurement framework that measures the long-term value of full-funnel content exploration while enabling fast experimentation. There are three layers in our solution: (1) a long-term “fresh content holdout” that serves as the north star for the cumulative value of content, (2) a content-level “graduation” corpus metric that functions as an intermediate proxy, and (3) a sensitive user-level short-term experiment metric measurable via user-segmented A/B experiments to facilitate rapid engineering iterations and launch decisions.

\begin{figure}
  \includegraphics[width=1.0\linewidth]{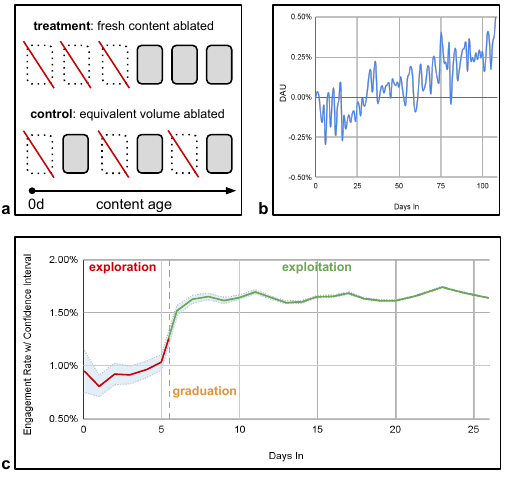}
    \caption{a)
    \textmd{Fair fresh content holdout experiment setup, holdout group (treatment) removes fresh content, production group (control) removes same amount of content randomly.}
    b)
    \textmd{The long-term fresh content holdout demonstrates that the short-term costs of exploration are outweighed by long-term retention benefits.}
    c)
    \textmd{Content graduation denotes the threshold at which an item's engagement value goes from uncertain to predictable.}
    }
    \Description{Three-panel figure showing the paper's measurement strategy: (a) the fair fresh-content holdout experiment setup, (b) long-term retention benefits outweighing short-term exploration costs, and (c) content graduation as the point where engagement becomes predictable.}
    \label{fig:measurement}
\end{figure}

We also evaluated the user-corpus codiverted experiment design \cite{wsdm24-explore} proposed previously. While it can address the treatment leakage issue when measuring content-level metrics, we found there are three major challenges  to this approach in a production setup. (1) We work on three major product surfaces across search and recommendations, and each surface has multiple candidate generators with different corpora (e.g., images, videos, products). It is cost-prohibitive to segment content into buckets globally across all these content corpora and product surfaces. (2) Since in such a setup when the experiment is at x\%, the results are only based on x\% of content corpus, it cannot accurately measure user-level metrics. However, in production it is critical to accurately measure the user-level metrics for content exploration improvements based on the full content corpus. (3) Limiting the content corpus to x\% will hurt user metrics and therefore limit the number of experiments that can be run in parallel, hurting experimentation velocity.

\subsection{\textbf{Fresh Content Value (North Star)}}
\label{sec:contentvalue}
The objective of full-funnel content exploration is to maximize the long-term user value generated by newly ingested content. We measure this via a long-term fresh content holdout that is applied consistently across all major discovery surfaces (Homefeed, Related Pins, and Search). In this design, the holdout cohort is limited to the content created before the start of the experiment, while the production cohort incorporates newly created content. To remove the inherent advantage of a larger candidate pool, we randomly remove an equivalent amount of content in the production group (based on content ID hash to ensure the same content is removed consistently for all requests), illustrated in Figure \ref{fig:measurement}.a. The observed engagement delta therefore quantifies the incremental benefit of the newly created and explored corpus (Figure \ref{fig:measurement}.b). Instead of demonstrating that a bigger explored content corpus leads to higher user engagement (already well established in \cite{wsdm24-explore}), our main goal is to ensure that the freshest explored corpus provides incremental value over the stale content corpus in the holdout (which has a sizable, fully-explored content corpus).

\subsection{\textbf{Content Graduation Corpus Metrics}}

While long-term holdouts provide the ground truth for fresh content value, they suffer from significant feedback latency. The cumulative benefits of full funnel exploration often persist well beyond the initial intervention window. Therefore, we define a content graduation metric as a robust intermediate proxy.
Our methodology is grounded in the observation that successful exploration transitions content from a high-variance "exploration" stage to a stable "exploitation" stage. Formally, we define the \textbf{content graduation} metric as the quantity (and proportion) of newly ingested content accumulating $X$ positive engagements within $Y$ days of ingestion, where we identified specific cumulative engagement threshold $X$ based our empirical data where the variance of engagement rate decreases significantly (Figure \ref{fig:measurement}.c).

In addition, differently from \cite{wsdm24-explore}, we want to avoid spending traffic on exploring content with low engagement potential. Therefore, we define the \textbf{explored low-engaging content} metric as the quantity of fully-explored content with the upper confidence bound of engagement rate less than some threshold $Z$. Combining both metrics, we then formally define \textbf{under-explored content} as content that is neither "graduated" nor "explored low-engaging".

\subsection{\textbf{Under-Explored Content Engagement Volume (Experiment Metrics) }}
As previous work \cite{wsdm24-explore} mentioned, user-segmented A/B experiments are not able to evaluate content-level metrics due to the shared corpus between arms. In our work, we use \textbf{under-explored content engagement}, defined as the volume of positive engagements received by under-explored content, in traditional user-segmented A/B experiments to measure system improvements.

There are several advantages of this experiment design: (1) this metric is strongly correlated with content graduation since more engagement to under-explored content is directly leading to graduation, (2) using positive engagement instead of impressions as the success criteria encourages exploration algorithms that have high relevance and low engagement tradeoffs, (3) we can confidently evaluate any overall user-level engagement tradeoffs when making launch decisions, and (4) since a standard user A/B setup is used, experimentation speed is maximized.

Theoretically, content level leakage is still possible in this setup, which causes under-estimation of the gain of this metric (i.e., better exploration in treatment could cause under-explored content to gain traffic in control). In reality, we observe that this is minimal, as (1) once the treatment accelerates content graduation, the graduated content will not be considered in the metric of the following day, and (2) the content a user sees in the control vs treatment group is typically distinct, especially when experiment traffic is low.

\begin{figure*}
  \includegraphics[width=1.0\linewidth]{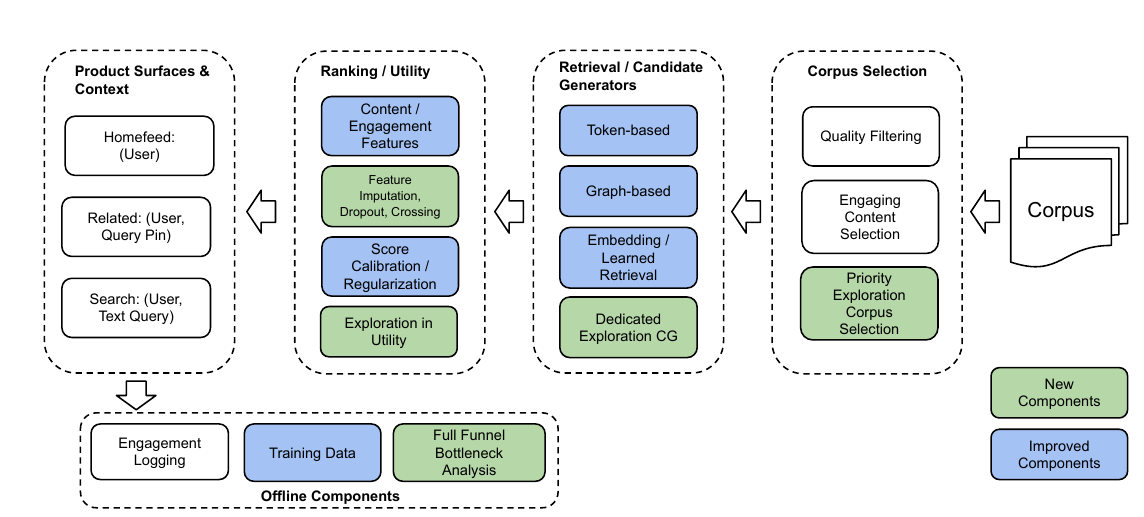}
    \caption{System Diagram: Full Funnel Content Exploration and Debiasing}
    \Description{A system diagram of a typical industry-scale multi-stage funnel at Pinterest for recommendation and search surfaces. The figure shows stages such as corpus selection, retrieval, ranking, and utility, and indicates how different product surfaces use different inputs, such as user information for homefeed and both user information and query context for search. It also shows where content exploration and debiasing are applied across the funnel.}
    \label{fig:funnel}
\end{figure*}

\section{System Overview}
\label{sec:system}
In this section, we discuss the high level system architecture: where we added new components and improved existing ones in the full funnel search and recommendation stack from end to end. We will also share specific insights and analysis that guided our development of this system.

\subsection{The Full-funnel Exploration and Debiasing Approach}
Figure \ref{fig:funnel} shows the diagram of a typical industry-scale multi-stage funnel for major recommendation and search product surfaces at Pinterest. Each product surface has its unique context as query input to the system: homefeed is a recommendation surface that only takes user information as input, while search takes the current query from the user to ensure relevance of the results in addition to user information for personalization.

To successfully introduce and gather user feedback for cold-start content, the entire content funnel requires stage-by-stage improvements. A significant bottleneck at any stage can throttle distribution to users. A key finding and contribution of this work is the recognition that cold-start content must effectively compete with existing content during the ranking and selection processes at every stage of the funnel and it is critical to reduce the system’s overall bias towards old content to ensure new content has a fair chance.

 The \textbf{corpus selection} stage is responsible for choosing the active corpus to index from the entire corpus by removing low quality and/or policy violating content and identifying high quality, engaging content to serve. Here we built new components to effectively select the most promising fresh content to explore since it is very costly, in both infrastructure and engagement, to fully explore every piece of new content at Pinterest's scale. The \textbf{retrieval stage} contains various types of candidate generators to retrieve the most relevant content given the context query to pass to the full ranking, e.g., based on text token like search query, graph random-walk \citep{eksombatchai2018pixie}, embedding or ML-based two tower architecture. Here we improved existing mechanisms and built dedicated candidate generators for cold-start content. In the \textbf{full ranking stage}, our main goal is to reduce the model's bias on both new and old content via improvements on features, architecture, loss function, calibration, as well as optimization in the final utility function for exploration. In the \textbf{offline components}, we improved training data sampling, and deployed a full funnel analysis to identify bottlenecks to guide our prioritization for further development.

\subsection{Full-funnel Bottleneck Analysis} \label{funnel-analysis}

The stages of a multi-layer recommendation funnel are highly interdependent and the realized impact of improvements at a given stage (e.g. ranking) can be constrained by the state of preceding stages (e.g. corpus selection and retrieval).  Consequently, it is critical to maintain a comprehensive view of bottlenecks and highest-ROI intervention points across the entire funnel. To this end, we employ a full-funnel log and bottleneck analysis that examines, for each stage, the input composition and output survival of cold-start versus existing content, specifically:

\begin{itemize}
    \item Input availability: fraction of input content that is under-explored and ingested within Y days;
    \item Output survival rate: probability of surviving in the top-k results among under-explored and newly ingested content.
\end{itemize}
Intuitively, a given stage can be bottlenecked, or at least inefficient, when fresh content is sufficiently present in the input but the survival rate in the output is low. Due to differences in product surface characteristics and the current state of underlying production systems, these inefficiencies can vary across surfaces, making this full funnel perspective critical to sequence work effectively.

\sectitle{Example insights from analysis:} On the Related Pins surface at certain time, we observed that the upper funnel (corpus and retrieval) was no longer the primary constraint since fresh content were returned at comparable rate as existing content. However, we found a significant drop occurs at the ranking stage: most of the fresh content were ranked below $k$ in the final output. This led us to shift focus to make ranking improvements.

On the Search surface during the same time frame, we have found the bottleneck starts at the retrieval stage. Given the strong relevance bar on search, we also hypothesized if there is enough relevant content for the query in the corpus. To be able to measure it, for a small sample of queries we over-retrieved a much larger portion of fresh content compared with the production system and ran the relevance model on the results; this validated the availability of relevant content at the corpus level and therefore confirmed that the bottleneck is at retrieval.
\section{Components}
\label{sec:components}
In this section we will discuss new or improved components added to each funnel stage in detail.

\subsection{Corpus Selection}

We have a large volume of Pins uploaded to Pinterest daily, making it infeasible to index and explore all of them. Therefore, we built a corpus selection pipeline that effectively identifies a  subset of high-potential fresh items to explore. By maintaining a dedicated exploration corpus and prioritizing the most promising fresh items, the system learns more quickly while reducing wasted traffic to content with low engagement potential.

\sectitle{Thompson Sampling:} Our first approach was to leverage Thompson sampling, a classic approach for explore / exploit. For each item, we sample the posterior engagement rate from $Beta(\alpha + e, n - e)$, where $e$ is the number of positive engagements, $n$ is the number of impressions and $\alpha$ is the prior based on overall engagement of the content provider. Then we pick the items with highest sampled score. In addition, we apply other business logic, such as ensuring diversity of content providers in the corpus.

\sectitle{Model-based Prior:} Over time, we would like to incorporate other signals into the prior, e.g., image quality, off-platform information about the content and provider. Naturally, we built a ML model to predict the empirical engagement rate of each item based on content and off-platform signals.

While detailed implementation of this model is not critical to our main finding, at a high level: for each item, the model predicts its engagement rate as $\alpha$. We then combine it with the empirical engagement rate to calculate the posterior, which is used for final selection:
\begin{equation}
\hat{r} \;=\; \frac{\alpha N + e}{N + n},
\end{equation}
where $e$ is the number of observed engagements, $n$ is the number of impressions, and $N$ controls the strength of the prior.

\sectitle{Selection:}
On each pipeline refresh, we select the top-$K$ items ranked by the posterior score $\hat{r}$. The exploration corpus is updated dynamically to align with our definition of ungraduated content: items are retired when 1) they have accumulated enough positive engagement to be graduated from exploration or 2) their upper confidence bound of engagement rate is lower than a threshold. This refresh-and-replace policy prioritizes exploration capacity for the most promising fresh content.

\subsection{Feature Improvements}
Having the right features is critical for debiasing the ML model, and in this section we will discuss the successful techniques we developed. Some of them are effective in improving the prediction on fresh content, while others can reduce the model's bias favoring existing old content. Most of these feature improvements can be applied to both retrieval stage model (e.g., two-tower learned retrieval) as well as the final ranking model.

\sectitle{Advanced Content Understanding Signals:}
Leveraging content-based features in cold-start is a well-studied topic \cite{sigir02-coldstart}. While content features were already widely leveraged in Pinterest's production system, we have made further improvements.
\begin{itemize}
    \item Content-only Embeddings: The existing system heavily leverages content embedding signals that are optimized for the pin-board graph or content co-engagement, like PinSage\cite{PinSage} and ItemSage\cite{ItemSage}. However, since a significant amount of fresh content is not well connected to the graph, relying on these embeddings leads to bias against fresh content. Therefore, we found it effective to incorporate additional content embeddings which are based on content only, such as Unified Visual Embedding \cite{UVEmbedding}, and, more recently, a new embedding leveraging large Visual-Language model (VLM) and CLIP visual text alignment (namely, PinCLIP\cite{PinCLIP}).
    \item Semantic ID: More recently, semantic id has been proposed as a novel technique for content representation  \cite{nips23-tiger}. Similar to content-only embeddings, we have identified that adding semantic id as an additional feature in existing models helps fresh content.
\end{itemize}

\sectitle{Feature Imputation:}
Generating advanced content embeddings often requires heavy ML inference, which means features will not be immediately available after content is created. Simply treating a missing feature with a naive default value can lead to a significant penalty for fresh content; we found simple feature imputation is effective to alleviate this issue. Specifically, we impute each dimension of the embedding by independently sampling from a normal distribution whose parameter is estimated based on the empirical distribution of that dimension in the content corpus.

\sectitle{Engagement Feature Dropout:} In traditional collaborative filtering (CF) based recommendation models, dropout \cite{volkovs2017dropoutnet} has been proposed as an effective technique for cold-start. In a typical industry scale setup like Pinterest, the ranking model is not traditional CF, but instead a deep neural net that has various features to represent users, items, and their interactions. The historical engagement counts and rates of items are often very predictive of future engagement, and therefore a naive model could be overfitting and overly dependent on these features. As a result, the model will tend to under-predict for fresh content and over-predict for old content.

To mitigate this issue, we apply dropout to historical item engagement features during training, forcing the model to generalize better and depend more on content signals. We evaluate two random dropout strategies:
(1) \textit{Uniform dropout}, where all item engagement features are dropped simultaneously.
(2) \textit{Individual dropout}, where each item engagement feature is dropped independently. Empirically, individual dropout consistently yields better performance. A feature-importance analysis confirms the effectiveness: models trained with individual dropout place less importance on engagement features and shift weight toward content and context based features. We do not find it to be effective to drop out user-level historical engagement features, therefore user-level personalization is not impacted by this.

\sectitle{Feature Crossing}: Feature dropout will only lead the model to depend less on engagement features globally, but in principle the ranking model should dynamically leverage the content features more for cold-start items and leverage item engagement history more to improve the prediction as the content gains more signals. To achieve such feature interaction in a deep neural net, we found it is effective to add content age signal, content features and historical engagement features to an explicit feature crossing module DCNv2 \cite{www-dcn}.

\sectitle{Off-platform Engagement Data:} Content may have existed long before it gets uploaded to Pinterest, and therefore an item new to Pinterest could have rich signals external to the platform that we can leverage. For example, for purchasable products, many merchants send Pinterest historical transaction data. Using these additional data sources to build strong priors on items and content providers (e.g., top sellers from each merchant, popular merchants) is very effective in improving fresh content performance.

\sectitle{Near Realtime Engagement Features:} If we can calculate the engagement rate of items in near real-time using data streaming frameworks like Flink, our models can react more quickly compared to daily batch pipelines. This is especially important when content is still in an early exploration stage. Therefore, we have built near realtime item engagement pipelines for fresh content while maintaining batch pipelines to reduce infrastructure cost for content that ages out.

\subsection{Retrieval}

At Pinterest, we use three different retrieval strategies: Token Retrieval, Graph Retrieval, and Embedding Retrieval. Token Retrieval is a term‑matching retrieval setup where documents are indexed into an inverted index and queries are expressed as structured query trees to fetch matching documents efficiently. Graph Retrieval \citep{eksombatchai2018pixie} performs random walks within a graph from the query node and returns the visited nodes (documents) as candidates. Embedding Retrieval is an ANN (Approximate Nearest Neighbor)-based search based on query vs document embedding similarity.

\sectitle{Token Retrieval:} As mentioned in Corpus Selection, we produce a dedicated corpus containing only unexplored content. For token retrieval, we built separate inverted index for the exploration corpus, such that during retrieval, we fetch a certain amount of candidates from this index, while the remaining candidates are retrieved from regular corpora containing fully explored, highly-engaging content. This dedicated exploration retrieval channel guaranties a stream of exploratory items in later stages.

\sectitle{Graph Retrieval:} Pixie is a family of Graph Candidate Generators that perform a random walk along a bipartite Pin-Board graph starting from query Pins (e.g., recently engaged Pins from the user) and return the visited Pin nodes as candidates \citep{eksombatchai2018pixie}. Such a graph naturally favors returning older content that has been saved to more boards since such Pins will have more connections to boards. In order to debias these candidate generators and return more unexplored content, we modify the graph traversal algorithm to use a weighted random walk where edges connected to unexplored Pins are assigned higher weights.

\sectitle{Embedding Retrieval:}
For embedding-based retrieval, we typically use two strategies: 1) based on pre-trained content embeddings that can present query and content, such as PinCLIP embedding \cite{PinCLIP}, 2) using the engagement data in the surface to train a two-tower network (one tower for query, one tower for content) \cite{covington2016deep}. Then for effective exploration, we have developed following approaches:
\begin{itemize}
    \item \textbf{Dedicated Exploration Channel}: We could build dedicated ANN index for the exploration corpus using pre-trained content embeddings. For learned retrieval, we could train a dedicated model only on exploration content by removing all engagement related features to minimize the bias (similar to \cite{wang2023fresh}).
    \item \textbf{Unified Learned Retrieval Model}: While training a dedicated model for exploration can minimize the bias, overtime it incurs significant overhead on infra and eng maintenance cost. Therefore, we developed unified learned retrieval models that handle both cold-start and existing content. By leveraging better features and debiasing techniques mentioned in the previous section, we observe this approach could effectively return enough fresh content while significantly reducing cost.
    \item \textbf{Debiasing the Query Tower}: In the Related surface, the query is (Pin, User, additional context). We have found in such case when the query is an existing Pin, the query-side features could introduce bias to the model to favor old content in the results. Therefore, we have made improvements to debias the query tower. Specifically, instead of crossing content and engagement features at deeper level of the neural networks, we first learn separate embeddings for various feature types (e.g., content features, engagement features), then add late stage fusion / crossing. This will ensure the final query embedding encodes more content features and therefore it could effectively reduce bias in the results.
\end{itemize}

\subsection{Final Ranking and Utility Layer}

As we move to the late stage of the search and recommendation funnels, we have a typical setup where one or multiple multi-task heavy ranking models with the most advanced architecture and features predict the various rewards of the items to the user (e.g., probability of a save, a click, or a relevance score); and a final utility layer that combines these rewards and other global factors, such as diversity, to form the final order of results to present to the user.

There are two key challenges at this stage. First, due to factors such as sparsity of cold start items in training data, ranking models often tend to under-predict rewards for fresh content and over-predict for old content. To solve this problem, in addition to the feature improvements discussed previously, we developed the following additional techniques to debias the model:

    \sectitle{Content-type-aware Calibration:} The calibration layer aligns model prediction scores with empirical action probabilities, allowing the final utility layer to accurately balance various rewards. However, naive global calibration cannot effectively reduce calibration issues that are specific to different content types, e.g., fresh content or other under-represented content types. To mitigate this problem, we consider important content categories as categorical signals in the calibration  model to ensure the model is well calibrated across all major content cohorts.

    \sectitle{Score Regularization and Training Data Augmentation}: We also introduced additional regularization terms in the loss function to encourage the models to align score distributions for fresh vs old content with similar content representations. To handle the potential different feature value distribution of cold start vs old items, we also developed techniques to generate augmented training data with mixed feature space. More details of this work can be found in \cite{ebrahimi2025warmer}. Note, due to unique characteristics of different product surfaces, the techniques focusing on training data augmentation or loss functions could require significant adjustments across surfaces. For example, the regularization improvement discussed here is originally developed on the Related Pins surface, but when we experimented on Search, we found that strong regularization caused significant pin-query relevance degradation and therefore need to be much more conservative.

After model prediction bias, the second major challenge in a large scale search and recommendation system is the feedback loop: the training data for predicting user engagement is collected from actual user engagement history, which depends on what the production system decides to show to the user in the first place. This is a well known issue, and in the context of this paper it means that if the production system severely under-predicts fresh content, the amount of training data on fresh content will be further reduced, and simple up-sampling or off-policy correction is not sufficient. To solve this problem, we developed several techniques to improve the final utility layer to allow more exploration of cold-start items:

\sectitle{UCB with Real-time Impressions:} Upper Confidence Bound (UCB) \citep{auer2002finite} is a classical approach to the explore-exploit tradeoff that applies an optimistic bonus to items based on the uncertainty in their value estimates. In a large scale system, accurately estimating variance or uncertainty is not trivial, so we apply a simple heuristic of using real-time impression counts to estimate uncertainty: fewer impressions mean higher uncertainty. Specifically:

    \begin{equation}
    \label{eq:heuristic_ucb}
    \text{UCB}_i = \frac{\alpha}{\sqrt{1 + \beta \cdot \text{impressions}_i}}
    \end{equation}

    where $\alpha$ controls the exploration strength and $\beta$ determines how quickly the bonus decays with accumulated impressions.
    
    We experimented with two integration approaches: (1) Adding the UCB term directly as a logit adjustment within the ranking model, and (2) incorporating it as an explicit term in the utility layer. Both implementations demonstrated strong performance in reducing selection bias and improving training data quality, though the utility layer integration offered greater flexibility for tuning independently of model retraining.
    
    \sectitle{Neural Linear UCB:} One issue with the impression-based UCB approach is that it ignores the feature space of content entirely. Neural Linear UCB \cite{neuralLinUCB} is a simple and scalable approach that leverages the last hidden layer of the neural network as a feature representation $\phi(x)$, derived from the raw model input x, with approximately linear relationship to the target reward: $y = \phi(x)^T\theta$. With that the LinUCB algorithm \cite{li2010contextual} can be used to compute the upper uncertainty bound for sample i:
    \begin{equation}
    \label{eq:nlb}
    \hat{y_i} =  \phi( x_i)^\top\theta + \alpha \sqrt{ \phi(x_i)^\top \Sigma \phi(x)},
    \end{equation}
where the covariance matrix $\Sigma = H^{-1}$ is derived from the design matrix:
    
    \begin{equation}
    \label{eq:designmat}
    H = \sum_{i=1}^{N}{\phi(x_i)\phi(x_i)^\top} + \lambda I.
    \end{equation}
    
    To integrate our solution, our ranking model training process consists of two steps: 1) large-scale training of the main model architecture using large volumes of sampled training data, 2) a lightweight calibration step using a smaller sample set, unbiased from sampling logic, to calibrate predictions against empirical engagement rates. We use a feature representation $\phi(x)$ learned during step (1) but update the covariance matrix with the latest data in step (2).
    
    To handle inherently binary tasks, such as click and save predictions, similarly to \cite{wsdm24-explore}, we avoid explicitly computing the ridge regression parameters $\theta$ but instead rely on the logit of the pretrained binary prediction as a cheap surrogate for the mean reward. Another practical issue we faced with this setup stems from the magnitude of the design matrix $H$ in Eq. \ref{eq:designmat} roughly scaling with the sample size $N$ and thus the posterior uncertainty tends proportional to $1/\sqrt{N}$. Since the sample size $N$ from step (2) can vary independently of the pre-training process in step (1) this risks unintentionally varying the overall magnitude of exploration from time to time. To overcome this we add a scaling term to the $\alpha$ parameter in Eq. \ref{eq:nlb}, $\alpha = \sqrt{\frac{N}{N_o}} \alpha_o$ where $N_o$ is set as a reference batch size to maintain stability.

    \sectitle{Role of Relevance in Search Utility:} The search surface has a much higher semantic relevance bar between the query and result. Therefore, the search stack has an advanced relevance model using mostly content-based features, which is an important contributor in the final utility in tandem with engagement prediction. Since the relevance model uses much more content-based features, its bias against fresh content is much lower than the engagement model. In addition, improvements to the relevance model will often also benefit fresh content. Based on this, we have done several specific optimizations on the search surface: 1) we increase the utility weight of relevance for fresh content to ensure highly relevant fresh content can be ranked highly despite the engagement model's under-prediction, 2) in UCB, we apply a minimum relevance threshold and scale the exploration term by relevance: $\text{UCB}_i' = \text{relevance}_i^\gamma \cdot \text{UCB}_i$ in order to ensure that low relevance content does not receive a large exploratory advantage.

\section{Experiments}
\label{label:exp}

\subsection{\textbf{Overall Impact: }}
We started to deploy and iteratively improve various parts of the overall system since 2024, and have delivered significant impact to user engagement and content eco-system health with specific side-wide metrics gains in Table \ref{exp-results-overall}.

\sectitle{Impact from Fresh Content Holdout:} As we explained previously, the fresh content holdout measures the incremental impact to site-wide user engagement (measured by our internal definition of successful sessions) from the freshly created and explored content - the north star goal of our system. As shown in Table \ref{exp-results-overall}, we have observed significant YoY relative growth of the session impact due to our system (note: 2025 vs 2024 is not capturing the full impact since there was significant growth 2024 vs 2023 as well). The impact is across priority markets in North America and international countries. We have also observed significant improvements in shopping sessions due to enhanced exploration of product content, particularly within North America, Pinterest's primary shopping market. Data from our 2025 content holdout reveals a total shopping session gain of 18.5\%, which is 5.3x higher than the overall, non-shopping session gains in North America, and 8.3x higher internationally. This suggests that shopping content distribution in particular is heavily reliant on newer items, highlighting the need for efficient exploration and debiasing. In fact, such content often faces more acute cold-start challenges due to its high volume of bulk uploads directly from merchant catalogs and a resulting lack of connectivity with the existing user-curated pin-board graph.

\begin{table}[ht]
\centering
\caption{Overall site-wide impact}
\label{exp-results-overall}
\begin{tabular}{ l  l }
\hline
\textbf{Metric}  & \textbf{Relative Lift} \\
\hline
Incremental Session Gain from Fresh Content \\
Holdout YoY 2025 vs 2024 & \\
- Overall Successful Session (North America) & +24\% \\
- Overall Successful Session (i18n) & +49\% \\
- Shopping Session (North America) & +63\% \\
- Shopping Session (i18n) & +29\% \\
\hline
Graduated Content Corpus (28-day) \\ YoY 2025 vs 2024 & +41\% \\
\hline
Cumulative Gain Site-wide Under-explored  \\ Content Engagement Volume & \\
- Homefeed Surface Holdout & +37\% \\
- Search Surface Holdout & +13\% \\
- Related Surface Holdout & +27\% \\
\hline
Number of Successful Content Providers  \\
YoY 2025 vs 2024 & +99\% \\
\hline
\end{tabular}
\end{table}

\sectitle{Graduation Corpus \& Under-explored Content Engagement Volume}: As shown in Table \ref{exp-results-overall}, we have observed significant YoY growth in the graduated content corpus (graduated within 28-day since creation), as well as significant gains in our experiment metrics (i.e., site-wide under-explored content engagement volume) across three major surfaces at Pinterest. The metrics reported are cumulative gains since H2 2024 from long-term A/B experiment holdout on each product surface. These metric gains have demonstrated the effectiveness of our overall system. In addition, it shows the practical effectiveness of our robust three-layer measurement framework: the under-explored engagement volume gain leads to the growth of graduated corpus, and then increased overall user engagement impact in fresh content holdout.

\sectitle{Content Provider Composition:}
Beyond delivering impact on user engagement, our system fosters a healthier content ecosystem. By reducing the "rich-get-richer" effect, the system allows more content providers to be more successful. We define the successful content providers as those with their content's site-wide engagement volume  share above a certain threshold, and we observed a 99\% year-over-year increase in this metric.

\begin{table}[h]
\centering
\caption{Consolidated component-level A/B experiment impact from the Related Pins surface.}
\label{exp-results}
\begin{tabular}{ l l l }
\hline
\textbf{Component}  & \multicolumn{2}{l}{\textbf{Engagement Volume Lift}}   \\
 & \textbf{All fresh} & \textbf{Underexplored} \\
\hline
Corpus Selection \\
and Retrieval & - & - \\
- Exploration Corpus & +7.75\% & +16.92\% \\
- Graph Retrieval & +3.91\% & +3.71\%\\
\hline
Feature improvements & - & - \\
- Retrieval & +2.48\% & +3.97\%\\
- Ranking & +5.57\% & +18.52\%\\
\hline
Model architecture   & - & - \\
- Retrieval  & +5.95\% & +5.52\%\\
- Ranking  & +8.63\% & +6.57\%\\
\hline
UCB exploration & - & - \\
- Impression-based Heuristic & +2.95\% & +3.12\%\\
- Neural Linear (over heuristic) & +5.83\% & +5.32\%\\
\hline
\end{tabular}
\end{table}

\subsection{Component-level Impact \& Insights:}
Throughout our work, we have launched a series of experiments that delivered statistically significant lifts in our target experiment metrics. To provide insights into the impact of the improvements in each component, we group these launches by the component they primarily target and summarize the results in Table \ref{exp-results}. Note, different from the overall impact gain in Table \ref{exp-results-overall} measured by surface-level long term holdout, the metrics here are simply aggregating the gains from individual launches since 2024. To make the comparison across components more meaningful, we pick a single recommendation surface (Related Pins) to report, and findings and conclusions from other product surfaces are directionally consistent as well. To provide additional insights on cold start improvement, we present engagement lift for both all fresh (created < N days) as well as under-explored content. All experiments we launched had strict guardrail on overall engagement tradeoff.

As evident from Table \ref{exp-results}, final metric lifts are observed across different components of the overall full-funnel system. However, introducing a dedicated fresh content exploration corpus for retrieval, together with ranking feature improvements (such as improved content embeddings) disproportionately improved under-explored performance. For UCB based exploration, we compare both the heuristic realtime impression-based method in Eq. \ref{eq:heuristic_ucb} and Neural Liner Bandit (NLB) in Eq. \ref{eq:nlb}. We observe that NLB achieves incremental gains over the heuristic approach. However, due to simplicity and easier interpretability of the heuristic, we choose that over the more complicated NLB in certain production scenarios. For example, on the Related Pins surface we deploy the heuristic approach in a light-weighted early funnel ranker and only use NLB in the most advanced final late stage ranking model.

Lastly, to achieve the final impact a key practical lesson from our development journey is that it is important to have the right sequencing and prioritization on which stage to focus on in a multi-stage production system. At high level, we started our effort at the corpus and retrieval stage, to ensure enough fresh content sent to the ranking stage and were able to realize significant impact even with a suboptimal ranking stage. Later we observed that the ranking stage bottleneck became much more significant via our full-funnel analysis, which led us to invest more on ranking improvements. Within the ranking stage, we also examine the availability of cold-start content in training data to prioritize feature / architecture improvements vs explicit UCB-exploration in utility.

\section{Related Work}
\label{sec:related}
\sectitle{Cold-start Content:} The content cold-start problems are extensively studied in traditional recommender system research, where collaborative filtering was the main approach. The most effective high level direction is adding more content-based matching into the model \cite{sigir02-coldstart, coldstart-survey}. Another research direction focuses on directly mitigating bias stemming from historical engagement features. Techniques like dropout on such historical features \cite{volkovs2017dropoutnet} offer a cost-effective solution but can introduce undesirable performance trade-offs. Specifically, sparse item ID embedding features in extremely large item spaces are prone to poor cold start performance. To address this ID sparsity, \cite{singh2024better} proposes leveraging Semantic IDs \cite{rajput2023recommender} as a compact, hierarchical representation of content semantics. Other strategies involve employing meta-learning to derive a content-based substitute for cold items \cite{luo2025online}. Additional approaches include transfer learning methods, which involve clustering items before model training and incorporating a loss term that facilitates the effective transfer of knowledge from warm items to cold items within each cluster \cite{chang2024cluster}.

\sectitle{Exploration Methods:} Multi-armed bandit approaches \citep{auer2002finite} are aiming to achieve the balance of explore vs exploit to optimize long term rewards in a dynamic system, and can be leveraged to mitigate the feedback loops in online recommender systems. A classic linear contextual Bandit approach LinUCB is proposed in \cite{li2010contextual}, later NeuralUCB \cite{neuralucb}, Neural Linear UCB \cite{neuralLinUCB} and Neural Thompson Sampling (NeuralTS) \cite{neuralthompson} extended the classic UCB or Thompson Sampling approaches in modern Deep Neural Networks (DNNs).

\sectitle{Industry-Scale Solutions on Cold-start:} Modern industry-scale recommender systems are generally multi-stage hybrid systems leveraging extensive content-based matching as well as historical user engagement. However, the challenges on content cold-start are still significant. On measurement, \cite{wang2023fresh,wsdm24-explore} proposed user-corpus co-diverted experiment design to prevent the corpus leakage issue when measuring the effect of exploration on content-level metrics. \cite{wsdm24-explore} also established the impact of larger discoverable corpus to long term user engagement. For algorithm improvements, \cite{wang2023fresh} proposed training dedicated two-tower retrieval model for cold-start content only, \cite{wsdm24-explore} proposed using Neural Thompson Sampling as explicit exploration mechanisms, \cite{coldstart-kuaishou} proposed an item to user mechanism to dynamically decide if a cold-start item should receive additional traffic and the relevant user cohort. Compared with these work, our new system innovates in measurement and every stage of the entire search / recommendation funnel with demonstrated effectiveness at industry-scale.

\section{Conclusion \& Future Work}
\label{sec:conclusion}
To summarize, in this paper we propose a new solution to solving the content cold-start problem in industry-scale search and recommender systems. Our innovations on the full-funnel approach, focus on debiasing all content, and robust measurement framework has demonstrated strong performance and delivered significant impact to the Pinterest content ecosystem. For future work, we plan to continue making improvements in the key components and incorporating more sophisticated objectives, such as long-term \textit{creator} value generated from exploration, into the system.

\begin{acks}
We thank Nivedita Bhaskhar, Joseph Bongo, Subrato Chakravorty, Yilin Chen, Pak Ming Cheung, Arkin Dharawat, Huizhong Duan, Weihao Gao, Rahul Goutam, Kurchi Subhra Hazra, Bella Huang, Jongho Kim, James Li, Liyao Lu, Sameed Qureshi, Dan Sedra, Eesha Shetty, Saurabh Wadwekar, Dylan Wang, Qi Wang, Weiguang Wang, Yi-Chin Wu, Jaewon Yang, Hao Zhang, Liang Zhang, Xianxing Zhang for their valuable contributions that made this work possible.
\end{acks}

\bibliographystyle{ACM-Reference-Format}
\bibliography{ref}


\begin{thebibliography}{26}


\ifx \showCODEN    \undefined \def \showCODEN     #1{\unskip}     \fi
\ifx \showISBNx    \undefined \def \showISBNx     #1{\unskip}     \fi
\ifx \showISBNxiii \undefined \def \showISBNxiii  #1{\unskip}     \fi
\ifx \showISSN     \undefined \def \showISSN      #1{\unskip}     \fi
\ifx \showLCCN     \undefined \def \showLCCN      #1{\unskip}     \fi
\ifx \shownote     \undefined \def \shownote      #1{#1}          \fi
\ifx \showarticletitle \undefined \def \showarticletitle #1{#1}   \fi
\ifx \showURL      \undefined \def \showURL       {\relax}        \fi
\providecommand\bibfield[2]{#2}
\providecommand\bibinfo[2]{#2}
\providecommand\natexlab[1]{#1}
\providecommand\showeprint[2][]{arXiv:#2}

\bibitem[Auer et~al\mbox{.}(2002)]%
        {auer2002finite}
\bibfield{author}{\bibinfo{person}{Peter Auer}, \bibinfo{person}{Nicol{\`o} Cesa-Bianchi}, {and} \bibinfo{person}{Paul Fischer}.} \bibinfo{year}{2002}\natexlab{}.
\newblock \showarticletitle{Finite-time analysis of the multiarmed bandit problem}.
\newblock \bibinfo{journal}{\emph{Machine learning}} \bibinfo{volume}{47}, \bibinfo{number}{2} (\bibinfo{year}{2002}), \bibinfo{pages}{235--256}.
\newblock


\bibitem[Baltescu et~al\mbox{.}(2022)]%
        {ItemSage}
\bibfield{author}{\bibinfo{person}{Paul Baltescu}, \bibinfo{person}{Haoyu Chen}, \bibinfo{person}{Nikil Pancha}, \bibinfo{person}{Andrew Zhai}, \bibinfo{person}{Jure Leskovec}, {and} \bibinfo{person}{Charles Rosenberg}.} \bibinfo{year}{2022}\natexlab{}.
\newblock \showarticletitle{ItemSage: Learning Product Embeddings for Shopping Recommendations at Pinterest}. In \bibinfo{booktitle}{\emph{Proceedings of the 28th ACM SIGKDD Conference on Knowledge Discovery and Data Mining}} (Washington DC, USA) \emph{(\bibinfo{series}{KDD '22})}. \bibinfo{publisher}{Association for Computing Machinery}, \bibinfo{address}{New York, NY, USA}, \bibinfo{pages}{2703–2711}.
\newblock
\showISBNx{9781450393850}
\href{https://doi.org/10.1145/3534678.3539170}{doi:\nolinkurl{10.1145/3534678.3539170}}


\bibitem[Beal et~al\mbox{.}(2026)]%
        {PinCLIP}
\bibfield{author}{\bibinfo{person}{Josh Beal}, \bibinfo{person}{Eric Kim}, \bibinfo{person}{Jinfeng Rao}, \bibinfo{person}{Rex Wu}, \bibinfo{person}{Dmitry Kislyuk}, {and} \bibinfo{person}{Charles Rosenberg}.} \bibinfo{year}{2026}\natexlab{}.
\newblock \bibinfo{title}{PinCLIP: Large-scale Foundational Multimodal Representation at Pinterest}.
\newblock
\showeprint[arxiv]{2603.03544}~[cs.CV]
\urldef\tempurl%
\url{https://arxiv.org/abs/2603.03544}
\showURL{%
\tempurl}


\bibitem[Chang et~al\mbox{.}(2024)]%
        {chang2024cluster}
\bibfield{author}{\bibinfo{person}{Bo Chang}, \bibinfo{person}{Changping Meng}, \bibinfo{person}{He Ma}, \bibinfo{person}{Shuo Chang}, \bibinfo{person}{Yang Gu}, \bibinfo{person}{Yajun Peng}, \bibinfo{person}{Jingchen Feng}, \bibinfo{person}{Yaping Zhang}, \bibinfo{person}{Shuchao Bi}, \bibinfo{person}{Ed~H. Chi}, {and} \bibinfo{person}{Minmin Chen}.} \bibinfo{year}{2024}\natexlab{}.
\newblock \showarticletitle{Cluster Anchor Regularization to Alleviate Popularity Bias in Recommender Systems}. In \bibinfo{booktitle}{\emph{Companion Proceedings of the ACM Web Conference 2024}} (Singapore, Singapore) \emph{(\bibinfo{series}{WWW '24})}. \bibinfo{publisher}{Association for Computing Machinery}, \bibinfo{address}{New York, NY, USA}, \bibinfo{pages}{151–160}.
\newblock
\showISBNx{9798400701726}
\href{https://doi.org/10.1145/3589335.3648312}{doi:\nolinkurl{10.1145/3589335.3648312}}


\bibitem[Chen et~al\mbox{.}(2025)]%
        {coldstart-kuaishou}
\bibfield{author}{\bibinfo{person}{Gaode Chen}, \bibinfo{person}{Ruina Sun}, \bibinfo{person}{Yinjie Jiang}, \bibinfo{person}{Tianxiang Li}, \bibinfo{person}{Yinlong Dai}, \bibinfo{person}{Qifan Shi}, \bibinfo{person}{Xiaojie Qin}, \bibinfo{person}{Jialong Fu}, \bibinfo{person}{Piaoyi Chen}, \bibinfo{person}{Ronggeng Huang}, \bibinfo{person}{Na Li}, \bibinfo{person}{Qi Zhang}, \bibinfo{person}{Jiang Liang}, \bibinfo{person}{Han Li}, {and} \bibinfo{person}{Kun Gai}.} \bibinfo{year}{2025}\natexlab{}.
\newblock \showarticletitle{A Cold-start Recommendation System at Kuaishou Designed from the Short-video Perspective}. In \bibinfo{booktitle}{\emph{Companion Proceedings of the ACM on Web Conference 2025}} (Sydney NSW, Australia) \emph{(\bibinfo{series}{WWW '25})}. \bibinfo{publisher}{Association for Computing Machinery}, \bibinfo{address}{New York, NY, USA}, \bibinfo{pages}{124–132}.
\newblock
\showISBNx{9798400713316}
\href{https://doi.org/10.1145/3701716.3715205}{doi:\nolinkurl{10.1145/3701716.3715205}}


\bibitem[Covington et~al\mbox{.}(2016)]%
        {covington2016deep}
\bibfield{author}{\bibinfo{person}{Paul Covington}, \bibinfo{person}{Jay Adams}, {and} \bibinfo{person}{Emre Sargin}.} \bibinfo{year}{2016}\natexlab{}.
\newblock \showarticletitle{Deep Neural Networks for YouTube Recommendations}. In \bibinfo{booktitle}{\emph{Proceedings of the 10th ACM Conference on Recommender Systems}} (Boston, Massachusetts, USA) \emph{(\bibinfo{series}{RecSys '16})}. \bibinfo{publisher}{Association for Computing Machinery}, \bibinfo{address}{New York, NY, USA}, \bibinfo{pages}{191–198}.
\newblock
\showISBNx{9781450340359}
\href{https://doi.org/10.1145/2959100.2959190}{doi:\nolinkurl{10.1145/2959100.2959190}}


\bibitem[Ebrahimi et~al\mbox{.}(2026)]%
        {ebrahimi2025warmer}
\bibfield{author}{\bibinfo{person}{Saeed Ebrahimi}, \bibinfo{person}{Weijie Jiang}, \bibinfo{person}{Jaewon Yang}, \bibinfo{person}{Olafur Gudmundsson}, \bibinfo{person}{Yucheng Tu}, {and} \bibinfo{person}{Huizhong Duan}.} \bibinfo{year}{2026}\natexlab{}.
\newblock \showarticletitle{Warmer for Less: A Cost-Efficient Strategy for Cold-Start Recommendations at Pinterest}. In \bibinfo{booktitle}{\emph{Proceedings of the ACM Web Conference 2026}} (United Arab Emirates) \emph{(\bibinfo{series}{WWW '26})}. \bibinfo{publisher}{Association for Computing Machinery}, \bibinfo{address}{New York, NY, USA}, \bibinfo{pages}{8105–8114}.
\newblock
\showISBNx{9798400723070}
\href{https://doi.org/10.1145/3774904.3792824}{doi:\nolinkurl{10.1145/3774904.3792824}}


\bibitem[Eksombatchai et~al\mbox{.}(2018)]%
        {eksombatchai2018pixie}
\bibfield{author}{\bibinfo{person}{Chantat Eksombatchai}, \bibinfo{person}{Pranav Jindal}, \bibinfo{person}{Jerry~Zitao Liu}, \bibinfo{person}{Yuchen Liu}, \bibinfo{person}{Rahul Sharma}, \bibinfo{person}{Charles Sugnet}, \bibinfo{person}{Mark Ulrich}, {and} \bibinfo{person}{Jure Leskovec}.} \bibinfo{year}{2018}\natexlab{}.
\newblock \showarticletitle{Pixie: A System for Recommending 3+ Billion Items to 200+ Million Users in Real-Time}. In \bibinfo{booktitle}{\emph{Proceedings of the 2018 World Wide Web Conference}} (Lyon, France) \emph{(\bibinfo{series}{WWW '18})}. \bibinfo{publisher}{International World Wide Web Conferences Steering Committee}, \bibinfo{address}{Republic and Canton of Geneva, CHE}, \bibinfo{pages}{1775–1784}.
\newblock
\showISBNx{9781450356398}
\href{https://doi.org/10.1145/3178876.3186183}{doi:\nolinkurl{10.1145/3178876.3186183}}


\bibitem[Li et~al\mbox{.}(2010)]%
        {li2010contextual}
\bibfield{author}{\bibinfo{person}{Lihong Li}, \bibinfo{person}{Wei Chu}, \bibinfo{person}{John Langford}, {and} \bibinfo{person}{Robert~E. Schapire}.} \bibinfo{year}{2010}\natexlab{}.
\newblock \showarticletitle{A contextual-bandit approach to personalized news article recommendation}. In \bibinfo{booktitle}{\emph{Proceedings of the 19th International Conference on World Wide Web}} (Raleigh, North Carolina, USA) \emph{(\bibinfo{series}{WWW '10})}. \bibinfo{publisher}{Association for Computing Machinery}, \bibinfo{address}{New York, NY, USA}, \bibinfo{pages}{661–670}.
\newblock
\showISBNx{9781605587998}
\href{https://doi.org/10.1145/1772690.1772758}{doi:\nolinkurl{10.1145/1772690.1772758}}


\bibitem[Luo et~al\mbox{.}(2025)]%
        {luo2025online}
\bibfield{author}{\bibinfo{person}{Yunze Luo}, \bibinfo{person}{Yuezihan Jiang}, \bibinfo{person}{Yinjie Jiang}, \bibinfo{person}{Gaode Chen}, \bibinfo{person}{Jingchi Wang}, \bibinfo{person}{Kaigui Bian}, \bibinfo{person}{Peiyi Li}, {and} \bibinfo{person}{Qi Zhang}.} \bibinfo{year}{2025}\natexlab{}.
\newblock \showarticletitle{Online Item Cold-Start Recommendation with Popularity-Aware Meta-Learning}. In \bibinfo{booktitle}{\emph{Proceedings of the 31st ACM SIGKDD Conference on Knowledge Discovery and Data Mining V.1}} (Toronto ON, Canada) \emph{(\bibinfo{series}{KDD '25})}. \bibinfo{publisher}{Association for Computing Machinery}, \bibinfo{address}{New York, NY, USA}, \bibinfo{pages}{927–937}.
\newblock
\showISBNx{9798400712456}
\href{https://doi.org/10.1145/3690624.3709336}{doi:\nolinkurl{10.1145/3690624.3709336}}


\bibitem[Pancha et~al\mbox{.}(2022)]%
        {PinnerFormer}
\bibfield{author}{\bibinfo{person}{Nikil Pancha}, \bibinfo{person}{Andrew Zhai}, \bibinfo{person}{Jure Leskovec}, {and} \bibinfo{person}{Charles Rosenberg}.} \bibinfo{year}{2022}\natexlab{}.
\newblock \showarticletitle{PinnerFormer: Sequence Modeling for User Representation at Pinterest}. In \bibinfo{booktitle}{\emph{Proceedings of the 28th ACM SIGKDD Conference on Knowledge Discovery and Data Mining}} (Washington DC, USA) \emph{(\bibinfo{series}{KDD '22})}. \bibinfo{publisher}{Association for Computing Machinery}, \bibinfo{address}{New York, NY, USA}, \bibinfo{pages}{3702–3712}.
\newblock
\showISBNx{9781450393850}
\href{https://doi.org/10.1145/3534678.3539156}{doi:\nolinkurl{10.1145/3534678.3539156}}


\bibitem[Panda and Ray(2022)]%
        {coldstart-survey}
\bibfield{author}{\bibinfo{person}{Deepak~Kumar Panda} {and} \bibinfo{person}{Sanjog Ray}.} \bibinfo{year}{2022}\natexlab{}.
\newblock \showarticletitle{Approaches and algorithms to mitigate cold start problems in recommender systems: a systematic literature review}.
\newblock \bibinfo{journal}{\emph{J. Intell. Inf. Syst.}} \bibinfo{volume}{59}, \bibinfo{number}{2} (\bibinfo{date}{Oct.} \bibinfo{year}{2022}), \bibinfo{pages}{341–366}.
\newblock
\showISSN{0925-9902}
\href{https://doi.org/10.1007/s10844-022-00698-5}{doi:\nolinkurl{10.1007/s10844-022-00698-5}}


\bibitem[Rajput et~al\mbox{.}(2023a)]%
        {rajput2023recommender}
\bibfield{author}{\bibinfo{person}{Shashank Rajput}, \bibinfo{person}{Nikhil Mehta}, \bibinfo{person}{Anima Singh}, \bibinfo{person}{Raghunandan Hulikal~Keshavan}, \bibinfo{person}{Trung Vu}, \bibinfo{person}{Lukasz Heldt}, \bibinfo{person}{Lichan Hong}, \bibinfo{person}{Yi Tay}, \bibinfo{person}{Vinh Tran}, \bibinfo{person}{Jonah Samost}, {et~al\mbox{.}}} \bibinfo{year}{2023}\natexlab{a}.
\newblock \showarticletitle{Recommender systems with generative retrieval}.
\newblock \bibinfo{journal}{\emph{Advances in Neural Information Processing Systems}}  \bibinfo{volume}{36} (\bibinfo{year}{2023}), \bibinfo{pages}{10299--10315}.
\newblock


\bibitem[Rajput et~al\mbox{.}(2023b)]%
        {nips23-tiger}
\bibfield{author}{\bibinfo{person}{Shashank Rajput}, \bibinfo{person}{Nikhil Mehta}, \bibinfo{person}{Anima Singh}, \bibinfo{person}{Raghunandan Keshavan}, \bibinfo{person}{Trung Vu}, \bibinfo{person}{Lukasz Heidt}, \bibinfo{person}{Lichan Hong}, \bibinfo{person}{Yi Tay}, \bibinfo{person}{Vinh~Q. Tran}, \bibinfo{person}{Jonah Samost}, \bibinfo{person}{Maciej Kula}, \bibinfo{person}{Ed~H. Chi}, {and} \bibinfo{person}{Maheswaran Sathiamoorthy}.} \bibinfo{year}{2023}\natexlab{b}.
\newblock \showarticletitle{Recommender systems with generative retrieval}. In \bibinfo{booktitle}{\emph{Proceedings of the 37th International Conference on Neural Information Processing Systems}} (New Orleans, LA, USA) \emph{(\bibinfo{series}{NIPS '23})}. \bibinfo{publisher}{Curran Associates Inc.}, \bibinfo{address}{Red Hook, NY, USA}, Article \bibinfo{articleno}{452}, \bibinfo{numpages}{17}~pages.
\newblock


\bibitem[Schein et~al\mbox{.}(2002)]%
        {sigir02-coldstart}
\bibfield{author}{\bibinfo{person}{Andrew~I. Schein}, \bibinfo{person}{Alexandrin Popescul}, \bibinfo{person}{Lyle~H. Ungar}, {and} \bibinfo{person}{David~M. Pennock}.} \bibinfo{year}{2002}\natexlab{}.
\newblock \showarticletitle{Methods and metrics for cold-start recommendations}. In \bibinfo{booktitle}{\emph{Proceedings of the 25th Annual International ACM SIGIR Conference on Research and Development in Information Retrieval}} (Tampere, Finland) \emph{(\bibinfo{series}{SIGIR '02})}. \bibinfo{publisher}{Association for Computing Machinery}, \bibinfo{address}{New York, NY, USA}, \bibinfo{pages}{253–260}.
\newblock
\showISBNx{1581135610}
\href{https://doi.org/10.1145/564376.564421}{doi:\nolinkurl{10.1145/564376.564421}}


\bibitem[Singh et~al\mbox{.}(2024)]%
        {singh2024better}
\bibfield{author}{\bibinfo{person}{Anima Singh}, \bibinfo{person}{Trung Vu}, \bibinfo{person}{Nikhil Mehta}, \bibinfo{person}{Raghunandan Keshavan}, \bibinfo{person}{Maheswaran Sathiamoorthy}, \bibinfo{person}{Yilin Zheng}, \bibinfo{person}{Lichan Hong}, \bibinfo{person}{Lukasz Heldt}, \bibinfo{person}{Li Wei}, \bibinfo{person}{Devansh Tandon}, \bibinfo{person}{Ed Chi}, {and} \bibinfo{person}{Xinyang Yi}.} \bibinfo{year}{2024}\natexlab{}.
\newblock \showarticletitle{Better Generalization with Semantic IDs: A Case Study in Ranking for Recommendations}. In \bibinfo{booktitle}{\emph{Proceedings of the 18th ACM Conference on Recommender Systems}} (Bari, Italy) \emph{(\bibinfo{series}{RecSys '24})}. \bibinfo{publisher}{Association for Computing Machinery}, \bibinfo{address}{New York, NY, USA}, \bibinfo{pages}{1039–1044}.
\newblock
\showISBNx{9798400705052}
\href{https://doi.org/10.1145/3640457.3688190}{doi:\nolinkurl{10.1145/3640457.3688190}}


\bibitem[Su et~al\mbox{.}(2024)]%
        {wsdm24-explore}
\bibfield{author}{\bibinfo{person}{Yi Su}, \bibinfo{person}{Xiangyu Wang}, \bibinfo{person}{Elaine~Ya Le}, \bibinfo{person}{Liang Liu}, \bibinfo{person}{Yuening Li}, \bibinfo{person}{Haokai Lu}, \bibinfo{person}{Benjamin Lipshitz}, \bibinfo{person}{Sriraj Badam}, \bibinfo{person}{Lukasz Heldt}, \bibinfo{person}{Shuchao Bi}, \bibinfo{person}{Ed~H. Chi}, \bibinfo{person}{Cristos Goodrow}, \bibinfo{person}{Su-Lin Wu}, \bibinfo{person}{Lexi Baugher}, {and} \bibinfo{person}{Minmin Chen}.} \bibinfo{year}{2024}\natexlab{}.
\newblock \showarticletitle{Long-Term Value of Exploration: Measurements, Findings and Algorithms}. In \bibinfo{booktitle}{\emph{Proceedings of the 17th ACM International Conference on Web Search and Data Mining}} (Merida, Mexico) \emph{(\bibinfo{series}{WSDM '24})}. \bibinfo{publisher}{Association for Computing Machinery}, \bibinfo{address}{New York, NY, USA}, \bibinfo{pages}{636–644}.
\newblock
\showISBNx{9798400703713}
\href{https://doi.org/10.1145/3616855.3635833}{doi:\nolinkurl{10.1145/3616855.3635833}}


\bibitem[Volkovs et~al\mbox{.}(2017)]%
        {volkovs2017dropoutnet}
\bibfield{author}{\bibinfo{person}{Maksims Volkovs}, \bibinfo{person}{Guangwei Yu}, {and} \bibinfo{person}{Tomi Poutanen}.} \bibinfo{year}{2017}\natexlab{}.
\newblock \showarticletitle{DropoutNet: addressing cold start in recommender systems}. In \bibinfo{booktitle}{\emph{Proceedings of the 31st International Conference on Neural Information Processing Systems}} (Long Beach, California, USA) \emph{(\bibinfo{series}{NIPS'17})}. \bibinfo{publisher}{Curran Associates Inc.}, \bibinfo{address}{Red Hook, NY, USA}, \bibinfo{pages}{4964–4973}.
\newblock
\showISBNx{9781510860964}


\bibitem[Wang et~al\mbox{.}(2023)]%
        {wang2023fresh}
\bibfield{author}{\bibinfo{person}{Jianling Wang}, \bibinfo{person}{Haokai Lu}, \bibinfo{person}{Sai Zhang}, \bibinfo{person}{Bart Locanthi}, \bibinfo{person}{Haoting Wang}, \bibinfo{person}{Dylan Greaves}, \bibinfo{person}{Benjamin Lipshitz}, \bibinfo{person}{Sriraj Badam}, \bibinfo{person}{Ed~H. Chi}, \bibinfo{person}{Cristos~J. Goodrow}, \bibinfo{person}{Su-Lin Wu}, \bibinfo{person}{Lexi Baugher}, {and} \bibinfo{person}{Minmin Chen}.} \bibinfo{year}{2023}\natexlab{}.
\newblock \showarticletitle{Fresh Content Needs More Attention: Multi-funnel Fresh Content Recommendation}. In \bibinfo{booktitle}{\emph{Proceedings of the 29th ACM SIGKDD Conference on Knowledge Discovery and Data Mining}} (Long Beach, CA, USA) \emph{(\bibinfo{series}{KDD '23})}. \bibinfo{publisher}{Association for Computing Machinery}, \bibinfo{address}{New York, NY, USA}, \bibinfo{pages}{5082–5091}.
\newblock
\showISBNx{9798400701030}
\href{https://doi.org/10.1145/3580305.3599826}{doi:\nolinkurl{10.1145/3580305.3599826}}


\bibitem[Wang et~al\mbox{.}(2021)]%
        {www-dcn}
\bibfield{author}{\bibinfo{person}{Ruoxi Wang}, \bibinfo{person}{Rakesh Shivanna}, \bibinfo{person}{Derek Cheng}, \bibinfo{person}{Sagar Jain}, \bibinfo{person}{Dong Lin}, \bibinfo{person}{Lichan Hong}, {and} \bibinfo{person}{Ed Chi}.} \bibinfo{year}{2021}\natexlab{}.
\newblock \showarticletitle{DCN V2: Improved Deep \& Cross Network and Practical Lessons for Web-scale Learning to Rank Systems}. In \bibinfo{booktitle}{\emph{Proceedings of the Web Conference 2021}} (Ljubljana, Slovenia) \emph{(\bibinfo{series}{WWW '21})}. \bibinfo{publisher}{Association for Computing Machinery}, \bibinfo{address}{New York, NY, USA}, \bibinfo{pages}{1785–1797}.
\newblock
\showISBNx{9781450383127}
\href{https://doi.org/10.1145/3442381.3450078}{doi:\nolinkurl{10.1145/3442381.3450078}}


\bibitem[Xia et~al\mbox{.}(2023)]%
        {TransAct}
\bibfield{author}{\bibinfo{person}{Xue Xia}, \bibinfo{person}{Pong Eksombatchai}, \bibinfo{person}{Nikil Pancha}, \bibinfo{person}{Dhruvil~Deven Badani}, \bibinfo{person}{Po-Wei Wang}, \bibinfo{person}{Neng Gu}, \bibinfo{person}{Saurabh~Vishwas Joshi}, \bibinfo{person}{Nazanin Farahpour}, \bibinfo{person}{Zhiyuan Zhang}, {and} \bibinfo{person}{Andrew Zhai}.} \bibinfo{year}{2023}\natexlab{}.
\newblock \showarticletitle{TransAct: Transformer-based Realtime User Action Model for Recommendation at Pinterest}. In \bibinfo{booktitle}{\emph{Proceedings of the 29th ACM SIGKDD Conference on Knowledge Discovery and Data Mining}} (Long Beach, CA, USA) \emph{(\bibinfo{series}{KDD '23})}. \bibinfo{publisher}{Association for Computing Machinery}, \bibinfo{address}{New York, NY, USA}, \bibinfo{pages}{5249–5259}.
\newblock
\showISBNx{9798400701030}
\href{https://doi.org/10.1145/3580305.3599918}{doi:\nolinkurl{10.1145/3580305.3599918}}


\bibitem[Xu et~al\mbox{.}(2020)]%
        {neuralLinUCB}
\bibfield{author}{\bibinfo{person}{Pan Xu}, \bibinfo{person}{Zheng Wen}, \bibinfo{person}{Handong Zhao}, {and} \bibinfo{person}{Quanquan Gu}.} \bibinfo{year}{2020}\natexlab{}.
\newblock \bibinfo{title}{Neural Contextual Bandits with Deep Representation and Shallow Exploration}.
\newblock
\showeprint[arxiv]{2012.01780}~[cs.LG]
\urldef\tempurl%
\url{https://arxiv.org/abs/2012.01780}
\showURL{%
\tempurl}


\bibitem[Ying et~al\mbox{.}(2018)]%
        {PinSage}
\bibfield{author}{\bibinfo{person}{Rex Ying}, \bibinfo{person}{Ruining He}, \bibinfo{person}{Kaifeng Chen}, \bibinfo{person}{Pong Eksombatchai}, \bibinfo{person}{William~L. Hamilton}, {and} \bibinfo{person}{Jure Leskovec}.} \bibinfo{year}{2018}\natexlab{}.
\newblock \showarticletitle{Graph Convolutional Neural Networks for Web-Scale Recommender Systems}. In \bibinfo{booktitle}{\emph{Proceedings of the 24th ACM SIGKDD International Conference on Knowledge Discovery \& Data Mining}} (London, United Kingdom) \emph{(\bibinfo{series}{KDD '18})}. \bibinfo{publisher}{Association for Computing Machinery}, \bibinfo{address}{New York, NY, USA}, \bibinfo{pages}{974–983}.
\newblock
\showISBNx{9781450355520}
\href{https://doi.org/10.1145/3219819.3219890}{doi:\nolinkurl{10.1145/3219819.3219890}}


\bibitem[Zhai et~al\mbox{.}(2019)]%
        {UVEmbedding}
\bibfield{author}{\bibinfo{person}{Andrew Zhai}, \bibinfo{person}{Hao-Yu Wu}, \bibinfo{person}{Eric Tzeng}, \bibinfo{person}{Dong~Huk Park}, {and} \bibinfo{person}{Charles Rosenberg}.} \bibinfo{year}{2019}\natexlab{}.
\newblock \showarticletitle{Learning a Unified Embedding for Visual Search at Pinterest}. In \bibinfo{booktitle}{\emph{Proceedings of the 25th ACM SIGKDD International Conference on Knowledge Discovery \& Data Mining}} (Anchorage, AK, USA) \emph{(\bibinfo{series}{KDD '19})}. \bibinfo{publisher}{Association for Computing Machinery}, \bibinfo{address}{New York, NY, USA}, \bibinfo{pages}{2412–2420}.
\newblock
\showISBNx{9781450362016}
\href{https://doi.org/10.1145/3292500.3330739}{doi:\nolinkurl{10.1145/3292500.3330739}}


\bibitem[Zhang et~al\mbox{.}(2021)]%
        {neuralthompson}
\bibfield{author}{\bibinfo{person}{Weitong Zhang}, \bibinfo{person}{Dongruo Zhou}, \bibinfo{person}{Lihong Li}, {and} \bibinfo{person}{Quanquan Gu}.} \bibinfo{year}{2021}\natexlab{}.
\newblock \bibinfo{title}{Neural Thompson Sampling}.
\newblock
\showeprint[arxiv]{2010.00827}~[cs.LG]
\urldef\tempurl%
\url{https://arxiv.org/abs/2010.00827}
\showURL{%
\tempurl}


\bibitem[Zhou et~al\mbox{.}(2020)]%
        {neuralucb}
\bibfield{author}{\bibinfo{person}{Dongruo Zhou}, \bibinfo{person}{Lihong Li}, {and} \bibinfo{person}{Quanquan Gu}.} \bibinfo{year}{2020}\natexlab{}.
\newblock \showarticletitle{Neural contextual bandits with UCB-based exploration}. In \bibinfo{booktitle}{\emph{Proceedings of the 37th International Conference on Machine Learning}} \emph{(\bibinfo{series}{ICML'20})}. \bibinfo{publisher}{JMLR.org}, \bibinfo{address}{Online}, Article \bibinfo{articleno}{1065}, \bibinfo{numpages}{11}~pages.
\newblock


\end{thebibliography}

\appendix

\end{document}